# Multidisciplinary analysis of pit craters at Hale crater, Mars

Mara Mantegazza[1], Mauro G. Spagnuolo[1], Angelo P. Rossi[2]

1. Instituto de Estudios Andinos, Departamento de Geología, FCEyN Universidad de Buenos Aires, Intendente Güiraldes 2160, C1428EHA Buenos Aires, Argentina
2. Constructor University, Campus Ring 1, 28759 Bremen, Germany.

Corresponding author: Mantegazza, Mara, mara.mantegazza1405@gmail.com or mmantegazza@gl.fcen.uba.ar

Abstract

Pit craters are circular to subcircular depressions that lack a rim and ejecta layer and typically have a conical shape. There are several mechanisms that can explain the formation of such depressions and they are associated with collapse due to the removal of subsurface material. Possible origins of pit craters include: volcanic processes (collapse of lava tubes, magmatic chambers, intrusion of dikes), karstic dissolution, extensional faulting or volatile processes.

In this work, we study pit craters in the north-estern external slope of Hale Crater. We identified more than 40 pit craters that present unique morphological features in association with different landforms associated with volatile activity in the area. We identified moraine like ridges, gullies and aprons together with pits that represent an interesting suite of landforms. After producing a detailed geomorphic map, we classified the pits into: i) single pit craters, ii) chain pit craters, iii) complex pit craters and iv) elongated pit craters.

We used DTMs to determine orientation and slopes of the pit craters-hosting terrains and to analyze elevation profiles of the different types of pits and their relation with the gullies and other features. In addition, we performed a spectral analysis using CRISM multispectral dataset in order to characterize the mineralogy of the sediments in the region and to identify any distinctive properties that could promote the formation of these depressions.

Here, we propose that pit craters are stratigraphically on top of the ice-related landforms and present complex relationships with the gullies. The spatial relationship between the pits and these structures, along with the absence of evidence of present or past volcanic activity and the lack of evidence of any extensional faulting allows us to propose that the origin of the pit craters in the study area might be related to some volatile process. We propose here that these particular pit craters at Hale crater, are morphologically similar to Icelandic depressions located in a glacial environment. We conclude that the landforms found in the area are in close relation with glacial or periglacial conditions and pit craters might be formed by sublimation/melting of ground ice.

1. Introduction

Pit craters are circular to subcircular and meter to kilometer size depressions. These features are characterized by a lack of ejecta layer and a raised rim. They are frequently observed on the Martian surface in many geologic settings and are associated with diverse collapse processes but also observed on other planets and moons (Wyrick et al., 2004). Proposed mechanisms for the formation of these depressions include: volcanic (Bardintzeff and McBirney, 2000; Mège et al., 2003), hydrologic/ice/karstic (Spencer and Fanale, 1990; Wilson and Head, 2002), structural/faulting (Banerdt et al., 1992; Ferrill and Morris, 2003; Kling et al., 2021; Mège et al., 2003; Tanaka and Golombek, 1989) and volatile processes (Lefort et al., 2009; Lefort et al., 2010; Morgenstern et al., 2007; Séjourné et al., 2011; Ulrich et al., 2010; Ulrich et al., 2012). A summary of the main mechanisms of pit crater formation are presented in Table 1 which is a modification of the overview presented by Wyrick et al. (2004). We investigate here the geologic context in which the studied pit craters are found to understand their origin.

In this work, we report the presence of pit craters located on the northeast external wall of Hale Crater. Hale Crater is located in the northeastern sector of the Argyre basin where Soare et al. (2017) describes the presence of depressions that resemble pit craters. They interpret that such depressions are generated by sublimation of ice-rich regolith or glacial-like icy mantle and classified by the authors as Type-1 and Type-2, respectively. However, the depressions described by those authors are typically polygonised, tear-dropped shaped and diameters ranging up to kilometers unlike the pit craters studied here (meter to hundred meter in size).

The main objective and scope of this work is to generate a detailed map of the area and the geomorphologic description of pit craters in the area. Then, we discuss the genesis of these pits in relation to a particular suite of glacial/periglacial landforms and gullies. Moreover, the area has been monitored by different cameras (HiRISE and CaSSIS) in relation to the recent activity of gullies. Considering these pit craters might be one of the younger features in the area, their study can provide insight into the recent history of the periglacial/glacial conditions in Hale Crater.

After a brief introduction to the Hale Crater geology, we present the methods and data used in section 2. The results of the morphological description of pit craters, together with the description of the geomorphic map are presented in section 3 and the topographic and spectral analysis in section 4. In the discussion we first discuss the morphology of pit craters in relation to the geology of the area followed by a discussion of possible terrestrial analogues and finally we explore potential origins for the pit craters studied here.

## 1.1. Regional setting: the geology of Hale Crater

Here we summarize the volatile related landforms of Hale crater as context for the map presented in section 3. Hale Crater is a large complex impact crater which is slightly elliptical in a northwest-southeast direction (125 km x 150 km) and possibly one of the most recent impact craters of its size (Jones et al., 2011).

The regional geology of Hale Crater is characterized by six units based on their location with respect to the crater rim (Jones et al., 2011). The central peak complex is formed by rocky outcrops aligned in the direction of the crater major axis and shows a topographic and shape asymmetry (the northern section is higher and wider). This asymmetry is also observed in the crater rim, with well-defined terraces and ridges in the northern part, while in the south the rim is poorly developed (Fig.1).

Hale Crater is surrounded by several channel systems that emanate from and transport material within distal regions of the proximal ejecta unit (He1). This unit was defined by Jones et al (2011) as a combination of ballistic and ground-hugging ejecta. The spatial relation between the channels with the ejecta layer and the transport of ejecta material indicates that fluvial activity started immediately after the impact. The liquid water required to form such fluvial systems most likely derives from the melting of subsurface ice during the Hale impact event (Jones et al., 2011; Tornabene et al., 2012). Based on crater counting, Jones et al. (2011) indicates a minimum age of 1Ga for Hale Crater, which would indicate fluvial activity in the early to middle Amazonian.

**Table 1**. Mechanisms associated with the formation of collapse depressions (Modified from de Wyrick et al., 2004).

| Proposed mechanism | | Description | Characteristics | References |
|---|---|---|---|---|
| Volcanic related processes | Lava tubes<br>Dikes<br>Collapsed Magma Chamber | Lava tube roof collapse, collapse after dike erupts in Plinian-type eruption with high dispersal rates collapse of elongate magma reservoirs. | Volcanic (basaltic) lava flows. Pit chains and troughs oriented down slope. Trace to magma source, likely in radial and/or concentric orientations. Evidence of reverse faulting along centerline, steep-walled or terraced pits. | Bardintzeff and McBirney (2000), Mège et al. (2003)) |
| Extension and dilation | Tension Fractures<br>Dilational Faulting | Tension fractures from extension; may be enhanced by weathering and erosion. Faulting along normal faults resulting in dilational segments. | Extensional stress field, opening of fractures, enlargement by weathering and erosion. | Banerdt et al. (1992), Ferrill and Morris (2003), Mège et al. (2003), Kling et al. (2021), Tanaka and Golombek (1989) |
| Karst Dissolution | | Chemical dissolution of soluble rock (e.g., carbonate). | Terrestrial karst-like topography controlled by dissolution, sinkholes. Evidence for an active hydrological cycle. Evidence for soluble rocks (e.g., carbonate). | Spencer and Fanale [1990] |
| Related to volatiles | Thermokarst, sublimation of ground ice | Sublimation of ground ice | Scalloped depressions commonly associated with polygons or small pits in the scarps. | Dundas (2017), Dundas et al. (2015), Lefort et al. (2010), Morgenstern et al. (2007), Séjourné et al. (2012), Soare et al. (2017), Ulrich et al. (2010), Viola et al. (2018) |

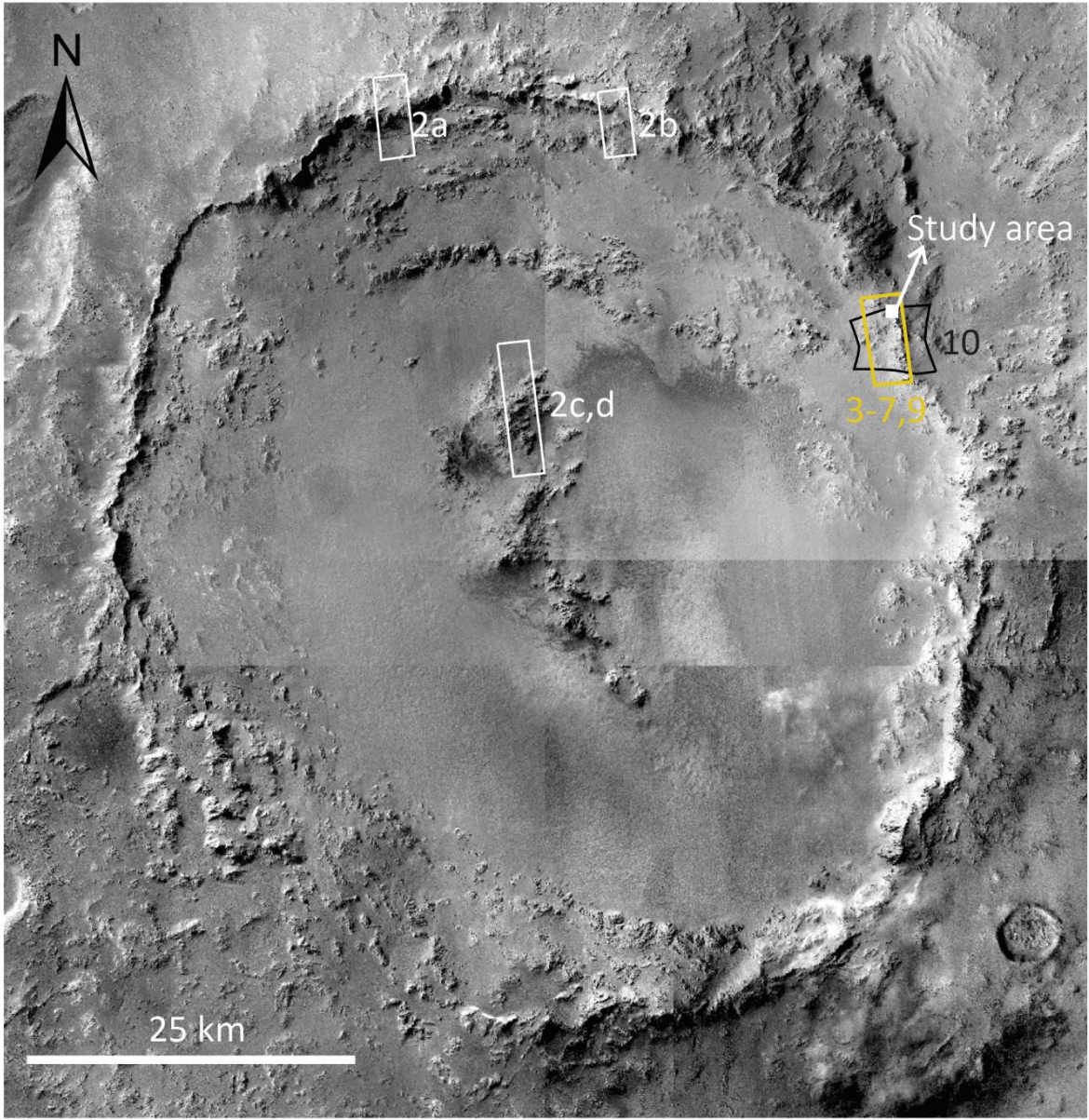

Figure 1: Mosaic of CTX images of Hale Crater (Murray Laboratory/MSSS/JPL). White boxes show the footprint of HiRISE images. Numbers correspond to figures with detailed observation. Black box shows the footprint of the CRISM image and the yellow box shows the footprint corresponding to the HiRISE image used for the mapping area.

The erosional activity and post-impact geomorphic processes might also be closely related with the volatiles contained in the regolith at and near the surface. Several authors have described channels, valleys, polygons and mounds in the ejecta layer of Hale Crater, both in the proximal and distal regions of the ejecta (Collins-May et al., 2020; El-Maarry et. al., 2013; Jones et al., 2011). These structures were related to volatile-rich material that was incorporated into the ejecta and accumulated all around the crater. Furthermore, El-Maarry et. al., (2013) proposes that the mounds and plateau fractures observed in the distal part of the ejecta are the result of post-impact glacial or periglacial activity. Below we focus on describing the landforms presented in the literature associated with volatile processes.

*Ice-related landforms on Hale Crater*

Typical landforms of glacial and periglacial conditions have been described in the mid-latitudes of Mars (see the comprehensive review by Souness and Hubbard, 2012). Several authors have described on the north inner wall of Hale Crater landforms commonly associated with sublimation or ice melting processes on Earth (e.g. gelifluction-like lobes, patterned ground and rock glaciers), resembling terrestrial periglacial environments (Hauber et al., 2011; Hiesinger and Head, 2002; Soare et al., 2019). Several models were proposed to explain the occurrence of glacial and periglacial conditions in mid-latitudes of Mars. Obliquity variations are thought to be the main factor controlling the preservation of $CO_2$ and $H_2O$ ices in Martian mid-latitudes (e.g. Mustard et al. 2001, Schorghofer 2007).

Hauber et al. (2011) described a protalus rampart on the inner north rim of Hale Crater. The front slope of this rampart is characterized by a well-defined polygonal surface and presents a few drop-shaped depressions (Fig. 2a). Polygons were described on other regions of the northern crater wall, along with gelifluction-like lobes and thermokarst depressions (Fig. 2b, Soare et al., 2019). Moreover, in the central peak, we identified moraine-like ridges that resemble lateral and terminal moraines on Earth (e. g. Arfstrom and Hartmann, 2005) and circular and rimless depressions along the base of the slope (Fig. 2c). In addition, we observed a sequence of talus cones on the pole-facing slope and at the bottom of these lobes there is a U-shaped valley (Fig 2d). This scenario with a lineated valley and lobate debris supplying the main valley are similar to a terrestrial glacial valley suite of features (e.g. Benn et al., 2003).

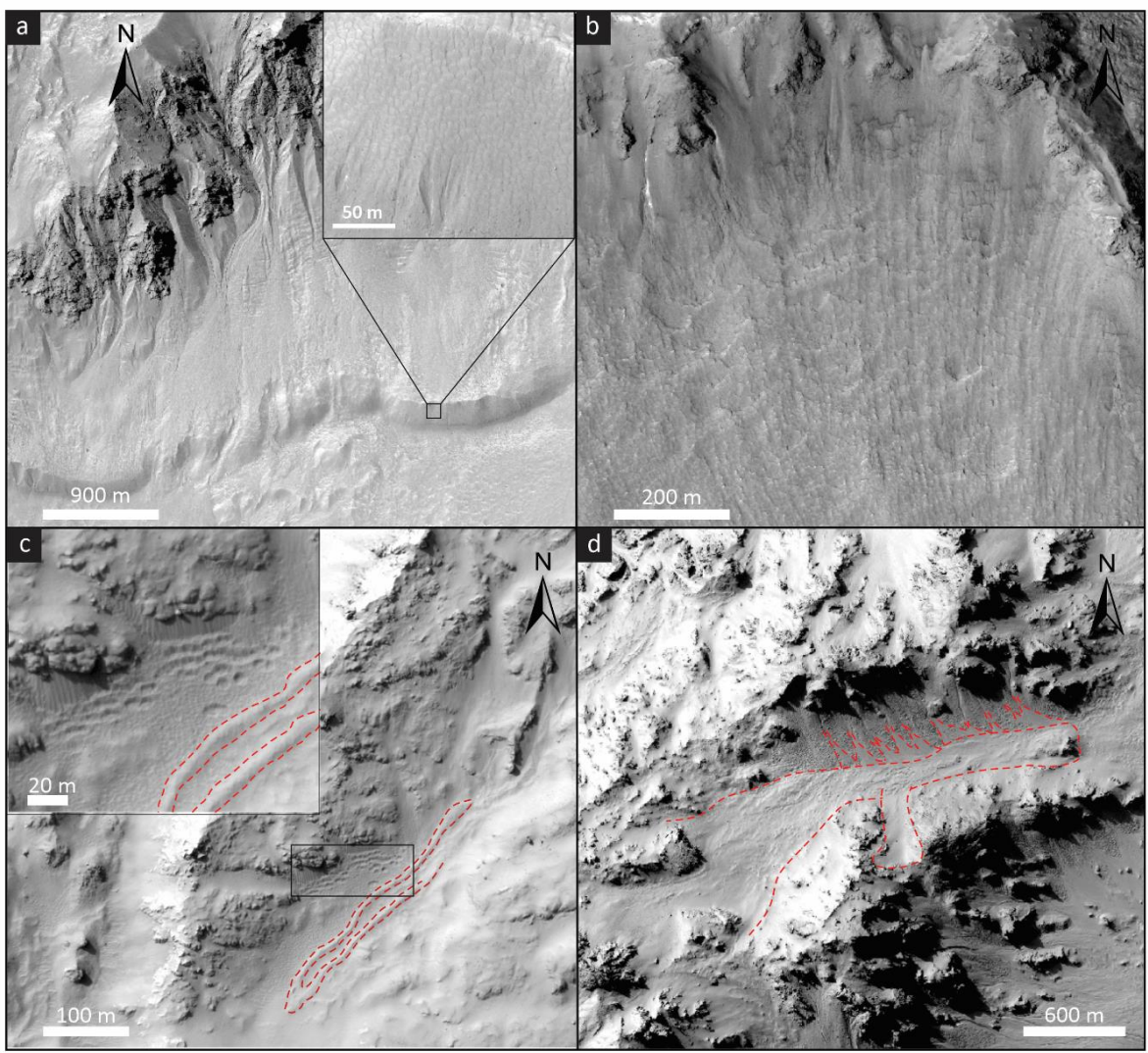

Figure 2: Ice-related landforms observed on Hale Crater. a) Protalus rampart with polygonal pattern on the front slope and drop-shape depression identified by Hauber et al., (2011) (detail of HiRISE image PSP_005688_1450). b) Thermokarst depressions identified by Soare et al. (2019) (near 35.6°S 36.15°W, detail of HiRISE image ESP_039392_1450). c) Terminal and lateral moraines related to single pits and pit crater chains in the central peak of Hale Crater (near 35.33°S 36.43°W). d) Glacier-like valley and talus cones (near 35.32°S 36.48°W). c) and d) are details of HiRISE image ESP_024346_1445. Red dashed lines outline the interpreted features.

*Recurring Slope Lineae*

The steep slopes of Hale Crater rim and central peak are the source of Recurring Slope Lineae (RSL) (Munaretto et al., 2020; Stillman & Grimm, 2018) and gullies (McEwen et al., 2007; Reiss et al., 2009; Soare et al., 2019). RSL show a distinct seasonal activity that can be explained by either wet (Huber et al., 2020; Levy et al., 2012; McEwen et al., 2011, 2014; Ojha et al., 2014,2015; Stillman et al., 2014), dry (Dundas, 2020; Dundas et al., 2017; McEwen et al., 2021; Munaretto et al., 2021; Munaretto et al., 2022; Schmidt et al., 2017; Schaefer et al., 2019; Vincendon et al., 2019) or hybrid processes (Bishop et al., 2021; Massé et al., 2016).

*Gullies*

There are abundant gullies and bright gully deposits (BGD) on the Hale Crater rim. Reiss et al. (2009) propose that in Hale Crater the distribution of gullies depends on the slope inclination and the availability of unconsolidated material. Kolb et al. (2010) models of BGD observed at Hale Crater indicate that they can be explained either by dry or wet flows. Although it has been described that in the past Hale Crater had surface aqueous activity considering the impact-associated channels, there is still debate if gullies are features related to water (e. g. de Haas et al., 2019)

*Ponded and pitted material*

The crater floor is characterized by a rough texture and ponded and pitted material is recognized (Jones et al., 2011). The pitted material is formed by the coalescence of quasi circular depressions (pits) of 10-30 meters in diameter with lack of rim and ejecta layer (Jones et al., 2011, Tornabene et al., 2012). The coalescence of the pits generates triangular intersections that lead to the formation of ridges. The result is a rough and irregular surface that appears as patches commonly on the crater floor and in the ejecta layers (Fig. 2). The pitted material is interpreted as collapse features related to volatile-rich impact melts (Boyce et al., 2012; Tornabene et al., 2012).

## 2.    Data and methods

For this study we reviewed 13 HIRISE images along the rim and interior of Hale crater (see appendix A) and from those images we could identify pit craters in only 3 of them. Moreover, of those three images only in two of them we found a close relation between pits and known volatile associated landforms. In the third image those pits are surrounded by ponded and pitted material and located near a region where Soare et al. (2018) described gelifluction and thermokarst features. We decided to analyze only one particular area as it had the most complete dataset available (topography, spectral data from CRISM and visible images HIRISE) and has a complete suite of superposing landforms that are been monitored due to the recent activity (gullies, moraine-like ridges, talus, pits).

We performed a detailed geological and geomorphological analysis on the eastern outer slope of Hale Crater and the dataset includes images provided by the Context Camera (CTX, Malin et al., 2007), the High Resolution Imaging Science Experiment (HiRISE, McEwen et al., 2007) and the Compact Reconnaissance Imaging Spectrometer for Mars (CRISM, Murchie et al., 2007). For analysis of the regional geology of Hale Crater and outlining the regions of interest, we used a ~5 m/pixel mosaic from Murray Lab (Dickson et al., 2018). The analysis of pit craters was conducted using a 0.25m/px HiRISE image (PSP_003209_1445_RED). Slope and aspect analyses were performed using a 1 m/px HiRISE DTM (DTEEC_002932_1445_003209_1445_U02) in Equirectangular projection (McEwen et al., 2007). Considering the extent of the HiRISE image the scale distortion factor along parallels for the whole scene ranges between ~1.001 and ~1.003, by using an Equirectangular projection (Statella, 2015). Since all our measurements are within the HiRISE image the scale factor is in the order of the pixel resolution. We also performed detailed topographic profiles of the studied features with a ~0.2m vertical precision (McEwen et al., 2007). Considering the scale of the features and the noise of the DTM, we smoothed it using a radius moving average of 10 meters to remove artifacts, like faceted areas (Munaretto et al., 2020; Munaretto et al., 2021; Schaefer et al., 2019). We use the corresponding 0.25m/px orthorectified HiRISE image (PSP_003209_1445_RED_A_01_ORTHO) for the measurements of the geometric properties of pit craters and the mapping.

For measurements of geometric and topographic parameters of pits we used ArcToolbox (Appendix B). The pit craters were mapped as ellipses and then we calculated the minor and major axis using the "*Minimum Bounding Geometry*" toolbox. With this tool we enclosed each ellipse with a rectangle, so the major and minor axis of each pit crater corresponds to the width and length, respectively, of the new polygon. To obtain slope and aspect measurement we used the "*Feature to Point*" tool to obtain the centroid of each pit crater. Finally, we extract the slope and aspect values of that centroid from the smoothed HiRISE DTM.

CRISM targeted observations are hyperspectral images that include full resolution targeted (FRT) and half resolution (HRS and HRL) products. In order to analyze the mineralogy of the study area, we used the FRT observations with a ~18 m/px spatial resolution. Since most of the pits have a sub-pixel scale of the CRISM resolution we did not attempt to obtain the mineralogy of individual features, instead we aim to characterize the regional geologic environment around the pit craters. We used a Map-Projected Targeted Reduced Data Record (MTRDR) product that includes the integration of the full spectral range (VNIR and NIR) with corrected I/F and atmospheric and photometric correction (Seelos, 2016). In addition, we used spectral summary parameters data that contains a suite of parameters used to highlight specific minerals (e.g. OLINDEX, HCPINDEX and LCPINDEX parameters for mafic mineralogy; Viviano et al., 2014). To process these images, we used the ENVI software and the CRISM Analysis Toolkit extension (Morgan et al., 2017).

In our research, we also compare martian pits with terrestrial pit craters located in Iceland and Argentina, to find analogs that allow us to understand the genesis of these structures and postulate a similar origin. For the terrestrial analysis we used data from Google Earth. This application offers free access to medium (Landsat, Copernicus) and high resolution images from WorldView-3 and WorldView-4 satellites (Maxar Technologies) with a resolution of 31 cm/px.

3. Geomorphic mapping

In this work, we analyzed a region located adjacent to the crater rim within unit He1 from Jones et al (2011) where we identify multiple pit craters (Fig. 3) in association with gullies and periglacial units. In the area we describe the different units according to morphological and textural characteristics and we produced a detailed geomorphic map.

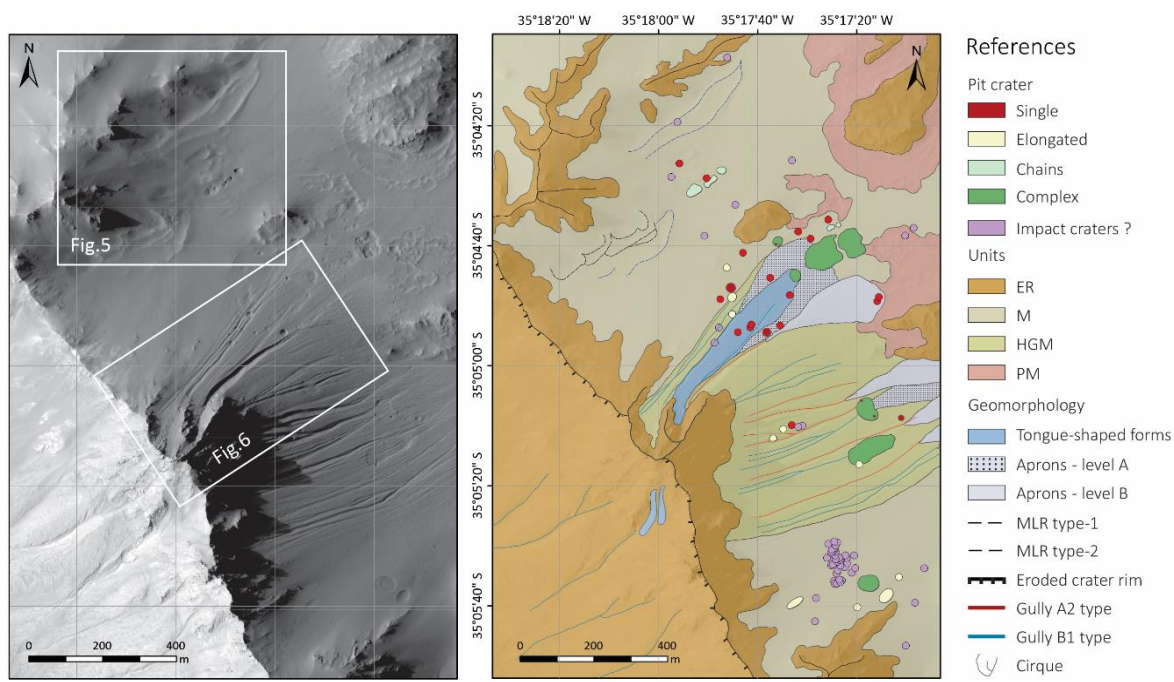

Figure 3: (Left) Details of HiRISE image (PSP_003209_1445_RED_A_01_ORTHO) showing the study area in an equirectangular projection. Boxes show the location of figures 5 and 6. (Right) Geomorphic map. Exposed rocks (ER); Mantling (M); Highly gullied mantling (HGM);

Pitted material (PM). The glacial and periglacial landforms and pitted material were mapped at a 1:3000 scale, while the pit craters were mapped at a 1:500 scale.

## 3.1. Mapping units

### 3.1.1. Pit craters

Pit craters are circular to subcircular depressions that, unlike impact craters, lack a raised rim or ejecta field associated with the impact. Pits occur as individual cone-shaped depressions or as pit crater chains (Wyrick et al., 2004). The pit craters differ from the pitted material described above, not only by their morphology and appearance, but also because of their extent. While pitted material is described as a rough surface texture, as shown in Fig.4, pit craters are discrete features with distinct circular to subcircular geometry. Moreover, unlike pit craters, the pitted material is formed by the coalescence of multiple pits resulting in linear ridges (Fig 7.c-f).

Impact craters are depressions commonly found on the surface of Mars and typically have a raised rim and an ejecta layer, however these features can be degraded as a result of post-impact processes (e.g Kress and Head, 2008; Mangold et al., 2012). To identify the pit craters we were especially attentive regarding whether or not a raised rim was visible. Multiple depressions have a well-defined rim (Fig 6a) while others have an elevated partial rim (Fig. 6b) highlighted by a shadow, in both cases no ejecta layer is associated. Moreover, some depressions are at the limit of the HiRISE image resolution making difficult to make a shadow analysis for the recognition of a raised rim. Considering the uncertainty in their classification we mapped all the depressions with a partial raised rim or with diameters smaller than 4 meters as "impact craters?" (Fig. 3).

In the study area we identified over 40 individual depressions with pit crater-like properties (Fig. 3, Appendix B). Depressions with a circular to subcircular morphology that lack a raised rim and ejecta layer were mapped as pit craters. We described the pits in terms of morphology, distribution, size and orientation and we classified them in four types.

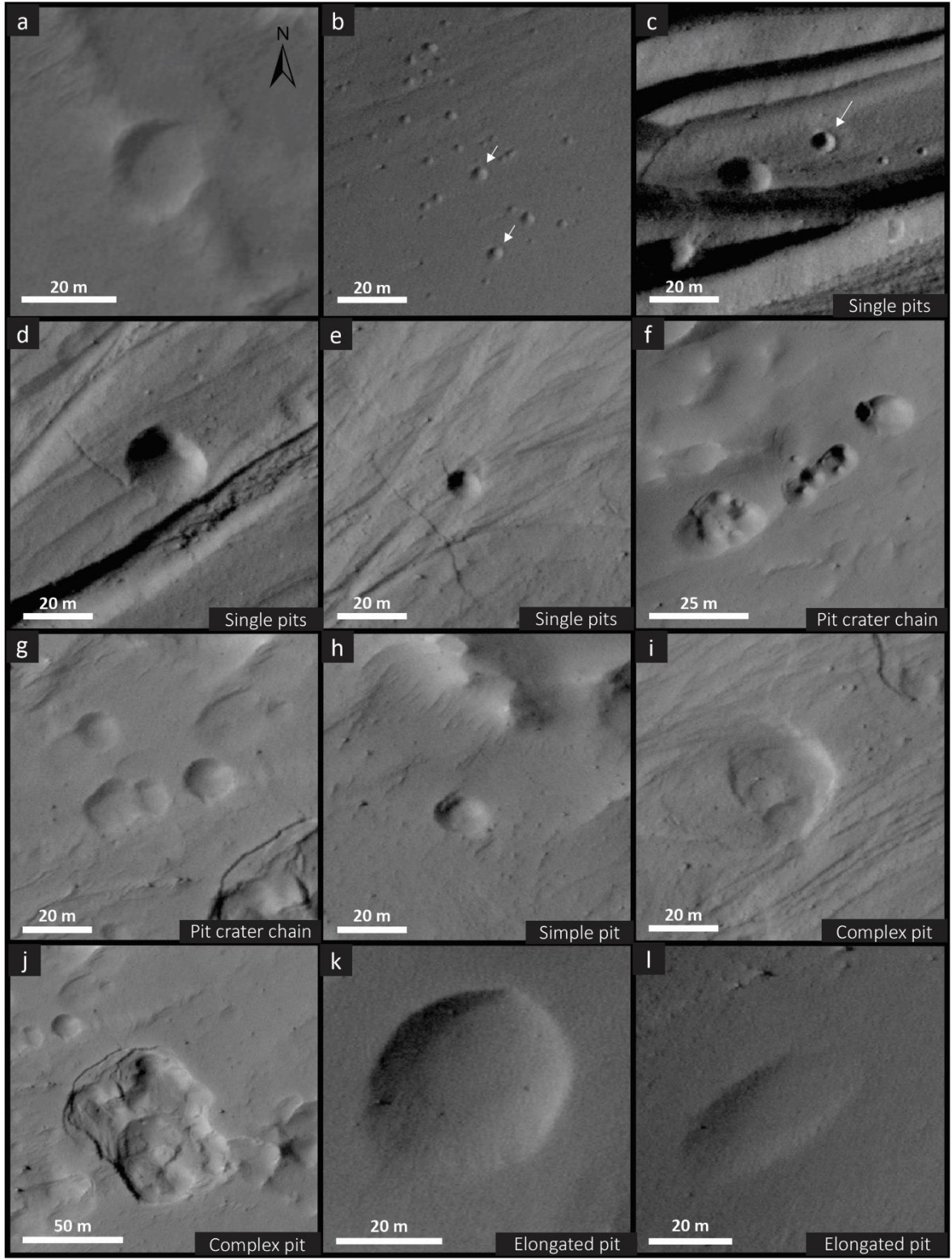

Figure 4: Detailed view of the morphologies of the depressions and their classification. a) Impact crater. b) Circular depressions with no ejecta associated, the white arrows point to the apparent rim. Pit crater morphologies: Single pits correspond to circular rimless depressions with a conic shape (c-e) or flat floors (g-h) and are usually dissected by gullies (c). Pit crater chains are formed by the alignment of single pit craters (f-g). Complex pit craters present an irregular geometry (i-j) and elongated pits correspond to single pits that present a ratio above 1:1.15 between the major and minor axis (k-l). Details of HiRISE image PSP_003209_1445_RED_A_01_ORTHO.

*Single pits*

Single pits represent the simplest kind of pit craters in the area. They have a circular to subcircular shape and the diameter ranges between 4 and 26 m. We identified two types of single pit, the ones with conic shapes (Fig. 4c-e) that are mostly found in the gullied mantling unit; and a second type with a flat floor (Fig. 4g-h). This type of single pits has diameters over 12 m, are not related to gullies and are located where the slope is <11°. We observed that the single pit craters seem to be clustered in the central region of the study area. To confirm this observation, we carried out a statistical spatial analysis comparing the arrangement of pits with a Poisson random distribution. The analysis verifies that single pits are distributed in clusters and not in a random manner (Appendix C).

*Pit crater chains*

Pit crater chains are linear features formed by the alignment of single pit craters in which their morphology remains well preserved. We identified two pit crater chains that present slightly distinct characteristics. The first corresponds to the coalescence of conical single pits and has a total length of 130 m (Fig 4f). In this case, a partial overlapping of the pits is observed, where the circular to subcircular geometry of the pits can still be recognized. In addition, some small pit craters (<5 m) can be found inside of the primary pit craters that make up the chain. The second type of pit crater chain has a linear arrangement of single flat-floored pits with a total length of 40 m (Fig. 4g). In either case, they are located in regions of low slope (Fig. 8d) and are characterized by the partial coalescence of pits.

*Complex pit craters*

In this class are grouped the pit craters with irregular geometry where one single pit cannot be recognized (Fig. 4i-j). Its morphology is far more complex than those previously described. In these cases, the pits show terrace-type structures possibly due to the coalescence of several single pit craters or several stages of collapse. In addition, in some cases, they show a highly irregular floor. Sizes vary from 30x40 m to 100x70 m and are located mainly in low slope areas (*see section 4.2*). These morphologies are distinct from the patches of pitted material found within the crater and in other regions of the ejecta layer (Fig. 2, Jones et al., 2011).

*Elongated pits*

This type of pit was classified considering the ratio between its major and minor axis. When this ratio was larger than 15% we mapped them as elongated. In the southern sector of the study area, we identified three pit craters with an elliptical geometry that falls under the 15% ratio criteria, with the maximum axis coincident with the direction of the slope. These are the shallowest pits compared with the other categories (Fig. 4 k-l) and are located in slopes from ~15 to ~45 degrees (Fig. 8d).

### 3.1.2. Moraine-like ridges

In the study area we observe two types of arcuate ridges that can be differentiated according to their relative position in the slope: (1) type-1: extensive ridges that extend along the slope and enclose a depressed area; and (2) type-2: curvilinear ridges that extends across the slope (Fig. 5). Type-1 extends over 200 and 300 meters (Fig. 5b-c) and can be traced up to the steeper slopes. Type-2 ridges correspond to curvilinear ridges that extend over 100 to 200 meters across the slope and overlay the type-1 arcuate ridges below (Fig. 5c). The distance between the steep slope and the ridges varies between 200 and 400 meters for the type-1 and type-2 ridges, respectively. These kinds of ridges have been described by several authors as moraine-like ridges (MLR) and have been commonly identified at the bottom of gullied eroded crater walls (Arfstrom and Hartmann, 2005). However, these structures have also been also identified in south-facing crater walls with no gullies (Berman et al., 2005). MLR resemble lateral and terminal moraines on Earth and have been related with ice-rich mantles (e. g. Arfstrom and Hartmann, 2005; Butcher et al., 2021; Hubbard et. al., 2014).

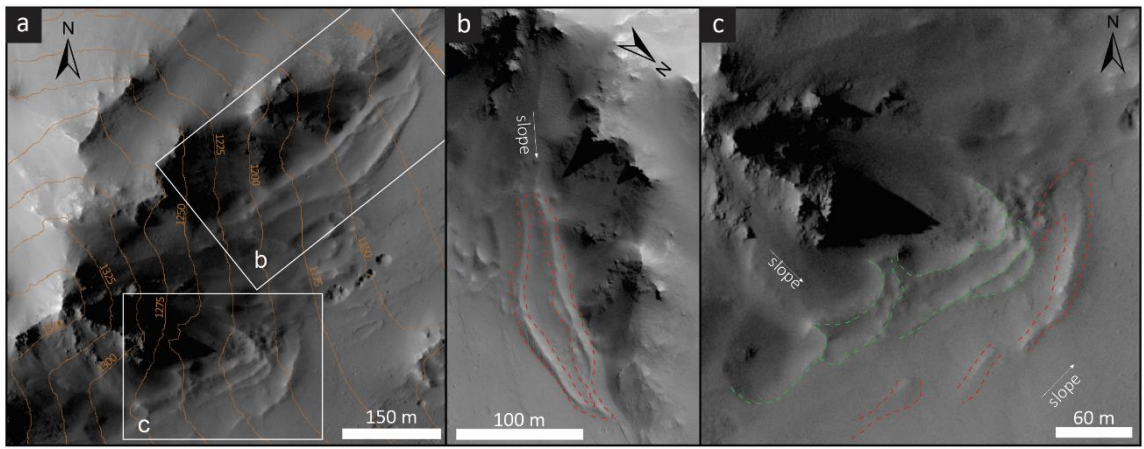

Figure 5: Moraine and moraine-like ridges. a) Local context of the landforms identified (center in 35.07°S 35.3°W). Boxes show location of the panels b and c. Contour lines with 25 m spacing obtained from the HiRISE DTM (DTEEC_002932_1445_003209_1445_U02). b) Elongated ridges enclosing a depressed central area could correspond to lateral and end moraines. c) In red, lateral moraine like features similar to those described in b. In Green, curvilinear ridges that extend across the slope and intersect a part of the lateral moraine below. Details of HiRISE image PSP_003209_1445_RED_A_01_ORTHO.

### 3.1.3. Gullies

Gullies are meter to kilometers scale erosional features characterized by an alcove-channel-apron morphology and have been associated with flows with a liquid component (Aston et al., 2011; Malin and Edgett, 2000; McEwen et al., 2007). However, several processes have been proposed for their formation, such as dry flows (Bart, 2007; Pelletier et al., 2008; Treiman, 2003) and processes related to ice melting (e.g. Conway et al., 2015; Costard et al., 2002). Moreover, since the observation of recent flows in gullies and the seasonal recurrence of the flows, $CO_2$ sublimation was proposed as one potential triggering mechanism for gully erosion (e. g. de Haas et al., 2019; Diniega et al., 2010; Dundas et al., 2010; Dundas et al., 2012). Gullies present a regional distribution between 30° and 60° latitudes and are mostly located on the polar-facing slopes (Balme et al., 2006; Conway et al., 2019; Harrison et al., 2015), a pattern also observed at Hale Crater (Raack et al., 2012).

In the study area, the gullies are located on northeast-facing slopes with inclinations between 35 and 27°. Gullies extend between 300 and 600 meters and we can identify two different morphologies. Some gullies exhibit a distinct V-shape profile and the typical alcove-channel-apron morphology cannot be identified. The alcove-channel section corresponds to a deep and carved main channel, sometimes branching into secondary channels that finishes in aprons. In most cases it is observed that the gullies only affect the mantling material and the bedrock is not involved in the erosional process (Fig. 6). Based on the classification scheme by Aston et al. (2011), we classified them as type A2 (mature) gullies. In the second type of gully, we recognized lengthened alcoves (Malin and Edgett, 2000) that end up in highly eroded aprons. In these cases, the erosion affects the bedrock and the topographic profiles show a wider channel section (Fig. 6). This type of gully was classified as type B1.

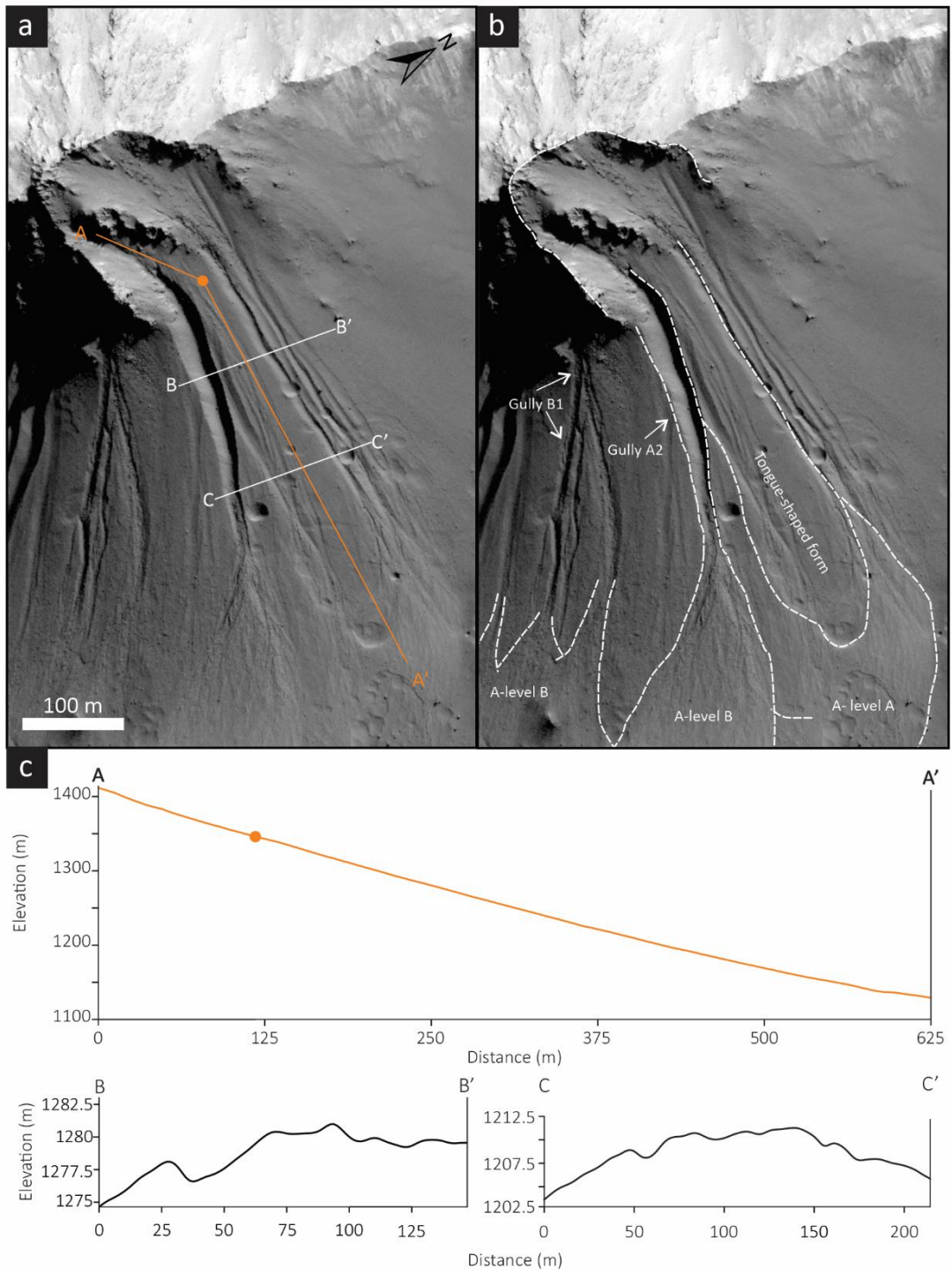

Figure 6: Examples of landforms identified on the east-facing slope a) Image without interpretation. In orange and white the locations of the topographic profiles displayed in c. b) Interpreted image: multiple lobes deposits and episodes of gully erosion incise the tongue-shaped form; the most recent episode corresponds to the V-Shape gully (type A2). Gullies type B1 ends with depositional lobes that corresponds to the aprons - level B. c) Elevation profiles of the tongue-shaped form. Detail of HiRISE image PSP_003209_1445_RED_A_01_ORTHO.

### 3.1.4. Lobate deposits

In the final section of some gullies lobate deposits emerge from the principal channel. These aprons exhibit an irregular texture that distinguishes them from the surrounding smooth mantle material. Aprons have a mean width/length ratio of 0.35 with width and lengths ranges between 35 and 165 m, 169 and 336 m, respectively. We identified several depositional aprons conforming to two levels of deposition: a stratigraphic lower level or Aprons level-A and an upper stratigraphic level or Aprons level – B. We define these levels based on the overlapping, the elevations profiles and the contacts between the deposits. Level-A aprons are related with type B1 gullies. The upper level of aprons is found in the final section of both types of gullies and in all cases the main channel continues into the apron itself. A representative example is shown in Figure 6, which shows a lobular deposit that that is not associated with any particular gully (apron level-A). Overlying this apron is another lobe-shaped deposit that can be traced out directly from a V-shaped gully that erodes the mantle material (gully A2 type). At the bottom of the slope, aprons are superimposed to the pitted material; Fig. 7e-f shows aprons overlapping the irregular texture of the pitted material.

On the east-facing slope we identify a tongue-shaped form that extends downslope for over 600 m originating from a cirque-like basin (Fig.6 a-b). The width of the landform ranges between 50 and 95 meters. The longitudinal topographic profile shows a slight concave profile with steeper slopes towards the head and homogeneous gentle downslope (Fig. 6a). The perpendicular topographic profiles exhibit a convex-up morphology (Fig. 6c). The limits of the landform are diffuse in the northern margin due to the presence of multiple gullies, so we outlined the lobe considering the breaks along the slope in an elevation profile.

### 3.1.5. Mantling

Mantling is defined as a cover of deposits showing a smooth and homogeneous texture (Fig. 3 and Fig. 7a-c). A few small impact craters (<20 m) are identified on this layer, evidenced by the rim, even though they lack an ejecta layer (Figs. 4a and 7c). This mantling extends downslope to the lower regions where it is in contact with the pitted material. At the top of the slope it is in contact with the rocky outcrops. As shown in Figure 3 on the hillslope there is a distinct highly eroded mantling patch that we classified as a separate unit (HGM).

### 3.1.6. Highly gullied mantling

This unit represents a highly eroded section of the mantle. It is characterized by the presence of multiple gullies cutting through the deposit layer. The limit between the mantle and the highly eroded mantle is more clearly defined in the northern sector of the unit and is limited by the northernmost gully. However, in the southern section the limit is diffuse due to a group of gullies covered by dust, indicating no recent activity. Moreover, this suggests that in the past the section of the mantle eroded by gullies extended southward (Fig. 7b). This unit is also detected in the analysis of CRISM hyperspectral images, showing a sharp spectral distinction between the mantling unit and the highly eroded mantle unit (see section 4.3). Some gullies exhibit a fan-shaped deposit (aprons) at the terminal section in transitional contact with the pitted material (Fig. 7e-f).

### 3.1.7. Pitted material

*Ponded and Pitted* material was described by Jones et al. (2011) as the coalescence of multiple circular to subcircular pits (Fig. 7d). In the geologic map presented by these authors, the Ponded and Pitted material unit is mapped only in the crater floor, although they mention the occurrence of this unit also in the ejecta layer. In the study area we identified pitted material that extends along the lower topographic regions that surround the rocky outcrops (Fig. 7e). The limits of this unit are clearly defined when it is in contact with the exposed rocks or mantling but the unit exhibits transitional contacts with gullies and terminal deposits, which seem to overlay the pitted material (Fig. 7f). Pitted material is differentiated from complex pits because it exhibits a widespread distribution over large areas and no individual closed depressions can be identified. Moreover, terrace-like morphology found in complex pits is not recognized in the pitted material unit.

### 3.1.8. Exposed rocks

The rocky outcrops form the crest of the Hale crater and the most elevated regions in the area (Fig. 7a and e). In the steep slopes we observed a clear or transitional contact with the mantling. Jones et al. (2011) interpret this unit as the bedrock prior to the origin of Hale Crater.

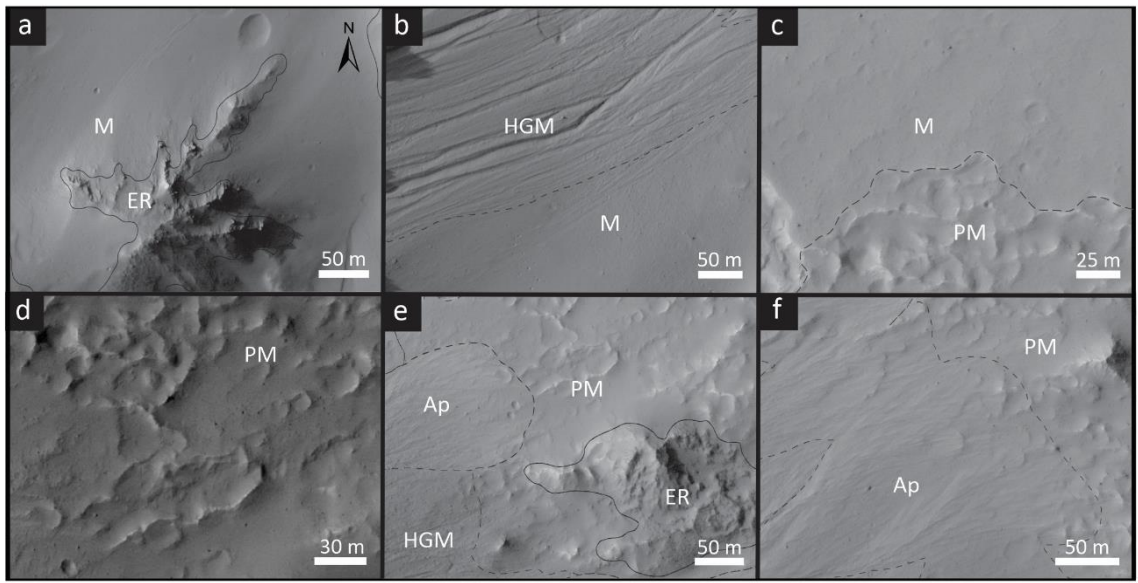

Figure 7: Examples of the mapped units. Dashed and solid lines correspond to transitional and well-defined contacts, respectively. Mantling (M); Highly gullied mantling (HGM); Exposed rocks (ER); Pitted material (PM); Aprons (Ap). Details of HiRISE image PSP_003209_1445_RED_A_01_ORTHO.

## 4. Topographic and Spectral analysis

### 4.1. Topographic analysis

In order to define the topographic properties of the region and investigate if there is a relation between the different pits with the slopes, we worked with a HiRISE DTM and calculated slope values and aspects. The regional slope decreases towards the north-east and most of the area has less than 45° slope except for some places close to the top of the rim of Hale. One of the main purposes of the slope analysis was to determine if there is a specific condition that promotes the occurrence of a particular type of pit crater. We tested the inclination and orientation of a random distribution to compare it with the patterns of the pits (Fig. 8). We simulated the distribution of 42 random points by restricting the area only where pit craters are located, avoiding, for example, the regions with exposed rock (see Appendix C).

We found that ~57% of the pit craters are located on slopes with an orientation between 55 and 75° with respect to north and in particular over half of the single pit craters occur in this range (Fig 8a). Single and elongated pits are located widespread on slopes of 12-28°, although elongated pits are restricted to elevation above ~1150 m. Most of the complex pits are restricted to slopes of 10-20° and the lowest areas (Fig. 8d), with the exception of one complex pit with an inclination of 26°. For the random distribution, the results show the regional northeastern orientation, and the slopes are higher in the most elevated areas following an almost linear trend between elevation value and slope (Fig. 8e).

In order to evaluate whether the samples of pit craters and random points are different, we performed a two-sample Kolmogorov-Smirnov test. We compared the distribution of aspect and slope value of the pit craters sample against the sample of random points. We obtained a p-value 0.0003 and 0.008 and a statistic of 0.45 and 0.35 for the aspect and slope distribution, respectively. Hence, the distribution of pit craters is statistically different from the random sample (Appendix C).

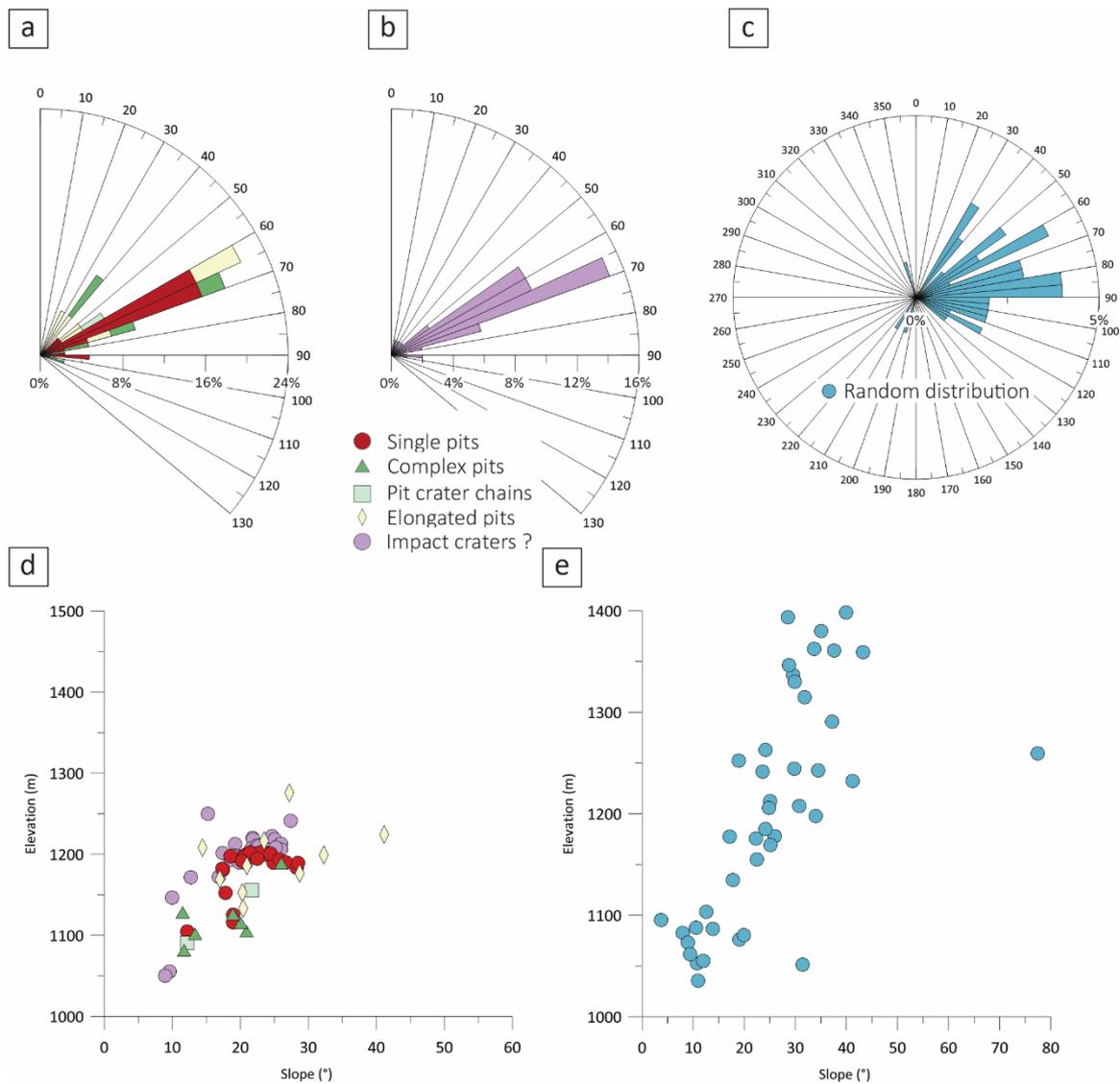

Figure 8: Topographic and slope distribution of the pit craters and a random sampling. a) Rose diagram of the slope orientation of all types of pit craters mapped. b) Same plot for the "impact craters?" sample. c) Same plot for a random sampling. d) Scatterplot of elevation vs slope value showing the distribution of mapped pit craters and possible impact craters. e) Same plot but for a random sampling. (Individual values are shown in table of Appendix B).

The small sample of complex pit craters and pit crater chains limits the statistical analysis, however we found that pit crater chains and complex pits are located only in regions with slopes less than 13° while single craters are distributed across all slopes.

In Figure 8b we show the aspect distribution of the depressions interpreted as possible impact craters. Over 75% of them are located in the slopes with an orientation between 50 and 75° respect to north. In terms of slope distribution, they are located widespread on slopes between 9 and 29° (Fig. 8d). In both orientation and inclination distribution, the "impact craters?" have a similar pattern to single pit craters. We tested the randomness of the sample of the possible impact craters comparing with a Poisson distribution and the results indicates a clustered distribution. We also performed a two-sample Kolmogorov-Smirnov test to compare the impact craters sample with simulated random points, the results verify that both are statistically different samples for the slope value and aspect.

Topographic cross sections show different pit crater geometries between the different pit types but in general they show cone shaped interior walls. Figure 9 shows examples of elongated, single, and complex pits; the topographic profiles were measured across the lower part of the pits and perpendicular to the regional slope. In the two cases of single pits, the profiles show a general depression due to gully erosion and a central cone-shaped depression. The gullies are identified as shoulders or minor depression in the profile (Fig. 9a-b). The cross section of the complex pit reveals the terrace-like structures and the irregular crater floor. The profile is strongly asymmetric and, in the north wall has a higher development of terraces while more steep slopes in the equatorial facing side (Fig. 9c).

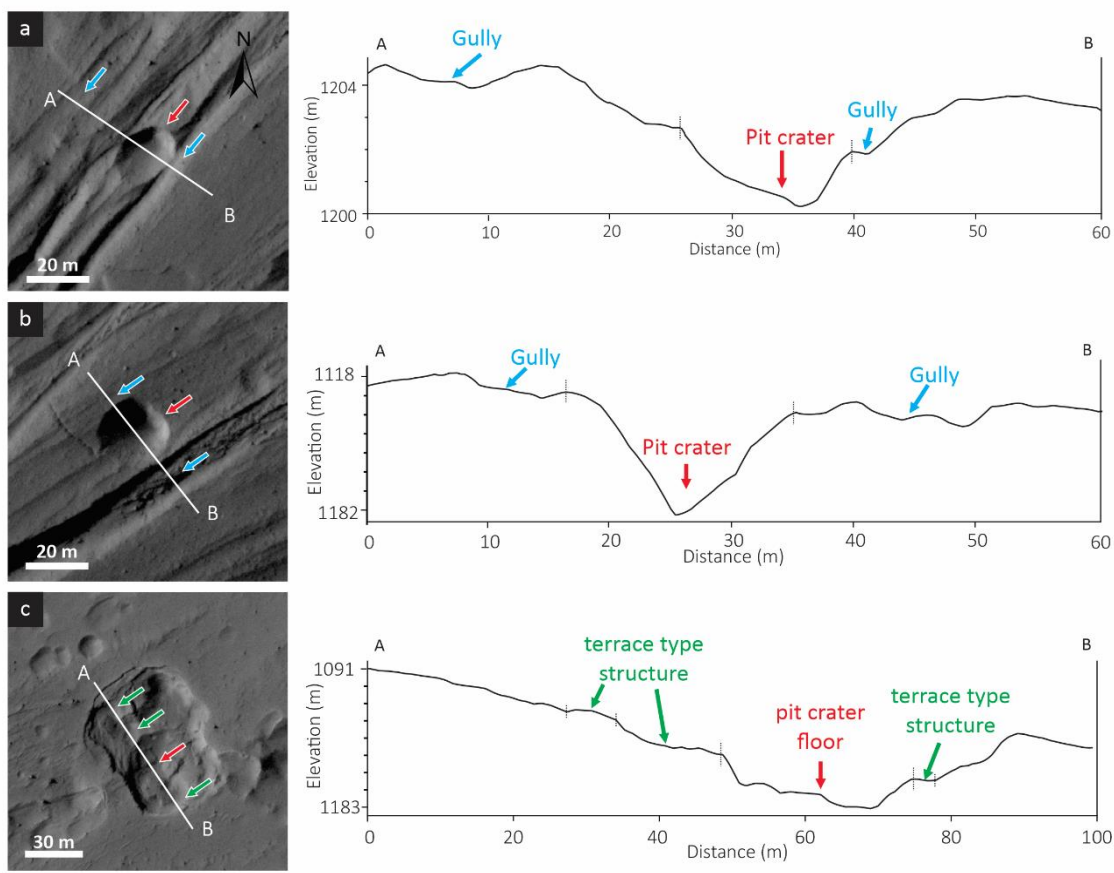

Figure 9: Elevation profiles for example of single and complex pit craters showing the principal geometry and the relation with gullies. a) an elongated pit located between the depression of the gullies, the vertical black lines represent the limits of the pit crater. b) Elevation profile of a single pit without gully dissection. c) Elevation profile for a complex pit showing the terrace-like structures, well defined on the north slope and faded on the southern slope. Elevation profiles are obtained from the filtered HiRISE DTM.

### 4.2. Spectral data

Considering the hypothesis that pits could be related to volatile processes, we looked for the presence of hydrated minerals such as sulfides, carbonates or phyllosilicates. We performed a spectral analysis that aimed to characterize the mineralogy of these areas and detect the presence of hydrated minerals.

Nuñez et al. (2016) conducted a comprehensive study of over 100 gullies at mid-latitudes on Mars and identified both gullies that exhibit distinctive characteristics compared to their surroundings and gullies that are spectrally indistinguishable from their surroundings. The authors analyzed an image that overlaps with our study area and found that gullies in the region exhibit a distinctive spectral characteristic. In most of the gullied slopes they identified pyroxenes as the main mineralogy while the surrounding zones are mainly olivine. These regions with a strong pyroxene spectral signature overlap with the pit craters studied here.

We examined different RGB combinations of several summary products from Viviano-Beck et al. (2014) to highlight the presence of hydrated minerals, however we did not find a conclusive mineralogical signature. A variation on mafic minerals can be detected in the MAF RGB composition where olivines or Fe-phyllosilicates are highlighted in red, pyroxenes with low-Ca and high-Ca content in green and blue, respectively (Fig. 10a).

Pyroxenes are a group of silicates with a variable composition of magnesium, iron and calcium and are detected by the presence of absorption bands centered at 0.9 and 1.8 µm, also the center of the bands depends on the calcium content and shifts to values of 1.05 and 2.3 µm in high-Ca pyroxenes (Mustard et al., 2005). Using the spectral parameters LCINDEX2 and HCPINDEX (for low calcium pyroxene and high calcium pyroxene, respectively) we selected regions of interest (ROI) to identify the mineral composition of the material. We calculated the ratio of the spectra for ROIs located in different positions on the gullied and pitted slope. For the ratio we divide the spectrum of each individual ROI by a nearby spectrally bland dark polygon. ROIs located in the first level of gullies deposition (aprons level-A) and in the tongue-shaped form exhibit absorption bands centered at 0.92 and 1.82 µm (Fig. 10b and c). This spectrum is compared with spectra from the CRISM resampled library and it is observed that the absorptions are consistent with low calcium pyroxenes pigeonite and estantite (Fig. 10c). Some pixels show long-wavelength band centers shifted to 0.98 and 2.05 µm microns with respect to the previously mentioned spectra, consistent with pyroxenes with a higher calcium content (Fig. 10d).

Most of the analyzed spectra present absorptions that are comparable with pyroxenes. However, we found that a few pixels corresponding to one of the deepest gullies channels present an absorption pair at 2.3 and 2.54 um (Fig. 10b and e). Although pyroxenes with high calcium content show absorptions at 2.3 µm, the absorption at 2.5 µm allows us to speculate with the presence of carbonates or siderite.

We identified two clearly distinct parts of the slopes that can be differentiated from the surroundings: one in the inner slope (outside the study area), and another in the outer slope where the pits are located (Fig. 10a). However, in the southern region where the gullies appear to have small deposits, and we infer that erosion dominates over deposition, there are no distinctive spectral properties from other areas in the image. These differences could be correlated with a difference in the activity of the gullies within the region, as also reported by Nuñez et al. (2016).

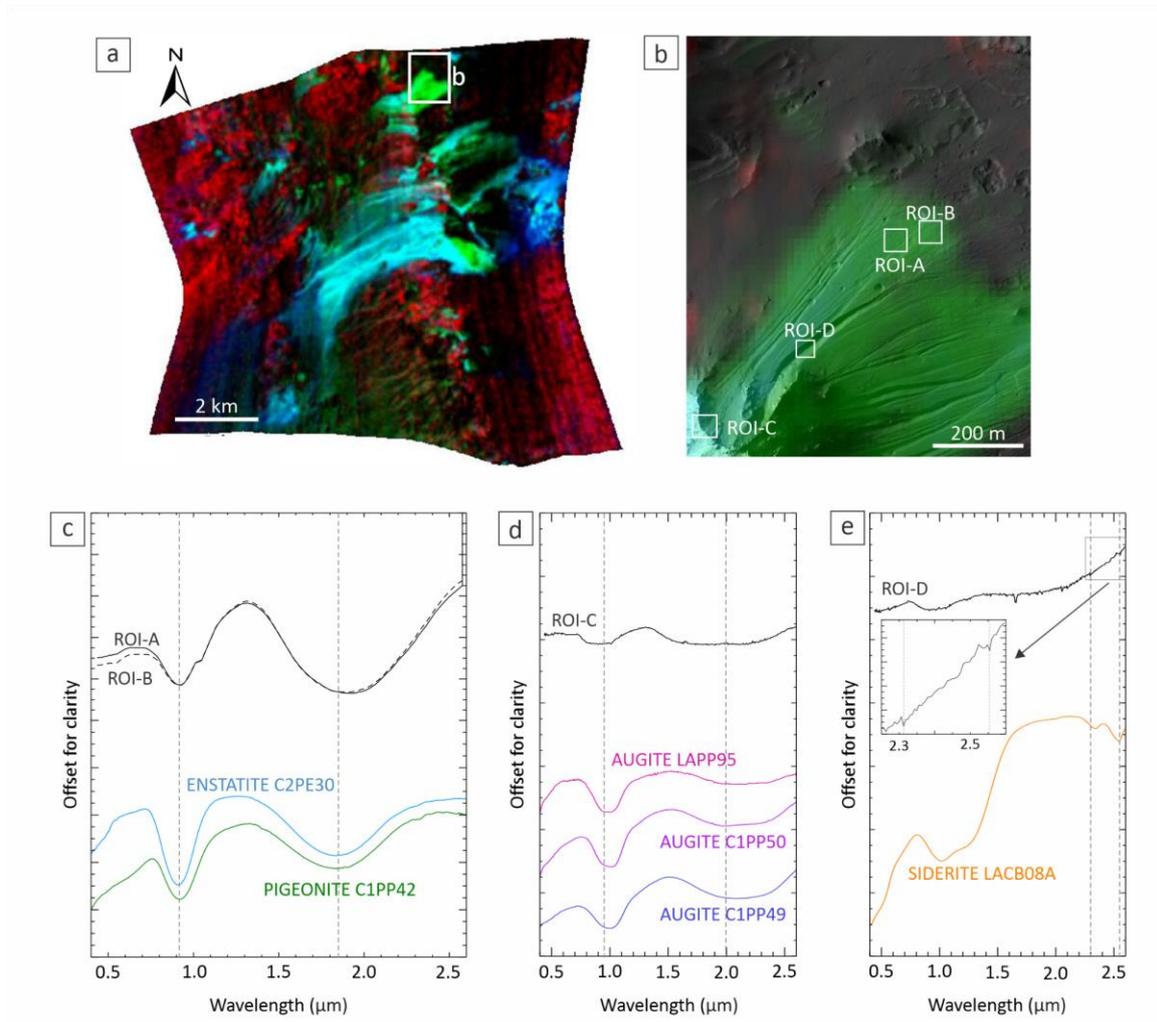

Figure 10: Results of the spectral analysis of the CRISM cube frt00004af7_07. a) RGB composite with summary product enhancing mafic minerals (R: OLINDEXX; G: LCPINDEX2; B: HCPINDEX2). The white box corresponds to the image b. b) RGB composite for mafic minerals over HiRISE of the study area. White boxes indicate the location of the pixels. c, d and e) Rationed spectrum from the ROIs shown in b, compared with spectrum from CRISM resampled library.

## 5. Discussion

### 5.1. Association of pits with surrounding landforms

The distribution of the pit craters and their stratigraphic relationship with the surrounding features may provide some clues regarding the origin and evolution of these depressions, as well as the history of the area. Pit craters are mainly distributed in the mantle deposits independent of the degree of erosion of the mantle material. In the more pristine mantle material, pit craters are generally smaller (<12m) and mostly single or elongated. The greatest variability in size and shape is associated with highly eroded mantle material and the lobate deposits.

The tongue-shaped form can be traced out from a cirque-like alcove which could be interpreted as a glacial-like form (e.g. Brough et al., 2019; Head et al., 2010; Hubbard et al., 2011; Souness et al. 2012). Moreover, the body shows a length/width ratio >1. Although, these elements are essential for the recognition of glacial-like forms according to Souness et al. (2012), certain elements characteristic of glacial forms are missing, such as evidence of down-slope flow like compressional/extensional ridges, crevassing or surface lineation. Williams et al. (2022) described the presence of rock glaciers – "like" landforms that lack the typical wrinkles perpendicular to flow direction on active terrestrial rock glaciers. The authors also identified higher slopes at their heads and lower in the distal sections and a convex morphology in a cross-section profile.

An alternative interpretation could be that they are alcove-apron type "gullies" described by Auld and Dickson (2016). However, the authors proposed that these landforms present a width/length ratio of 0.2±0.05, while the lobate form described here has a ratio of ~0.12. In addition, this lobate form shows a smoother texture that differs from the other aprons identified in the region.

Despite the tongue-shaped form having some of the characteristic elements of a rock glacier, the shape and ratio, the fact that it can be traced out directly from a cirque and the convex upward morphology in cross sections (Fig. 6); there is not sufficient evidence to classify it as a glacier feature. Subsurface studies might help to clarify whether this deposit is a mix of rock and ice.

Considering that some pits are found above the MLR and the aprons, and also the link with the gullies, we speculate a genetic origin to these features. Moreover, the single pit craters located in the central peak of Hale Crater are also related with MLR and lobate features, so this could indicate a relation between pit craters and volatile activity (Fig. 2c). During the Amazonian Period oscillation in orbital parameters lead to the migration of ice to lower latitudes on Mars resulting in glacial and periglacial environments at those latitudes (Dickson et al., 2012; Head et al., 2003). These oscillations can explain the occurrence of glacier-like forms, gullies and moraine like ridges as suggested by Arfstrom and Hartmann (2005). Moreover, Soare et al. (2018) proposed that the periglacial complex found in the northern internal wall of Hale could be the result of climate conditions during the Late Amazonian.

Suggesting the origin of the previously described ice-related landforms is straightforward, since the climatic conditions that explain their formation are better understood. However, as mentioned early in this work, several models describe gully formation with no relation with water/ice processes. Given the spatial relationship between the gullies and the pits, and the fact that the pits seem to occur sometimes before and sometimes after the gullies, we suggest that the genesis of both forms is somehow related. Assuming that gullies were generated by liquid-water debris flows involving processes such as shallow aquifer discharge (e.g. Goldspiel and Squyres, 2011; Hartman et al.,2003; Malin and Edgett, 2000; Mellon and Phillips, 2001) or subsurface snow/ice melt (e.g. Christensen, 2003; Costard et al., 2002; Gilmore and Phillips, 2002; Reiss et al., 2011), we would expect the formation of hydrated minerals. Although the spectral analysis results show no evidence of hydrated minerals, Heldmann et al. (2010) reports that the lack of hydrated mineral signature is not enough to rule out a wet origin for gullies and hence, also the pit craters. In addition, if water was present in the formation of the gullies, this may lead to dissolution of certain materials such as carbonates or salts.

During the last decade, based on the geographic relation of gullies and $CO_2$ reservoirs, several authors support the theory that gullies are driven by $CO_2$-related processes (e. g. de Haas et al., 2019; Diniega et al., 2010; Dundas et al., 2010; Dundas et al., 2012; Pasquon et al., 2016). In addition, de Haas et al. (2019) conducted the analysis of gullies located near the study area of this work and revealed that recent activity in those gullies could be triggered by small proportions of $CO_2$ ice.

Pasquon et al. (2016) describe gullies over linear dunes that often end up in a single pit or in a group of them. Several hypotheses are proposed to explain the occurrence and distribution of these terminal pits such as ice sublimation or $CO_2$ gas escaping. However, the spatial relation between gullies and pit craters observed on Hale Crater has no similarity to the arrangement described for linear dune gullies. Instead, in this work, we observed pit craters located at different positions with respect to gullies and moraines deposits. This might imply a common scenario for all these structures but not necessarily a genetic relation.

5.2.    Terrestrial Analogues

On Earth there are different types of depressions and some of them have similar characteristics to the pit craters we have studied on Mars. Considering the possible presence of carbonates and the association with surrounding ice-related landforms, we analyzed pit craters of two possible terrestrial analogs that present similar characteristics with the pits identified in Hale Crater.

Our first analogue is found in the Puna area of northwestern Argentina; we identified an alluvial fan with pit-like depressions. We identified single pit craters with diameters between 80 and 190 meters as well as a chain of pit craters with a total length of 700 meters (Fig. 11a-c). These structures are found in an alluvial fan located close to and even overlaying the Hombre Muerto Salar where Vinante and Alonso (2006) describe the presence of carbonate facies (Fig. 11a). The presence of carbonate facies at the southern edge and probably below the fan suggests that these pits might have formed by collapse of the loose material over the void produced by the dissolution of carbonates below the alluvial fan. The identification of a carbonate in the upper section of one of the martian gullies presents a similar scenario for the pits on Hale Crater where lower dissolution of material might be one of the possible processes forming the pit craters. Carbonate deposits on or near the surface of Mars and the formation of landforms resulting from karstic dissolution have been reported at several locations on Mars (Grindrod and Balme, 2010; Jackson et al., 2011; Wezel and Baioni, 2014). At Iani Chaos Sefton-Nash et al. (2012) identified spectral signatures characteristic of kieserite and gypsum. In the same region, Baioni and Tramontana (2015) described the morphology of pit crater-like depressions and proposed a karstic formation process.

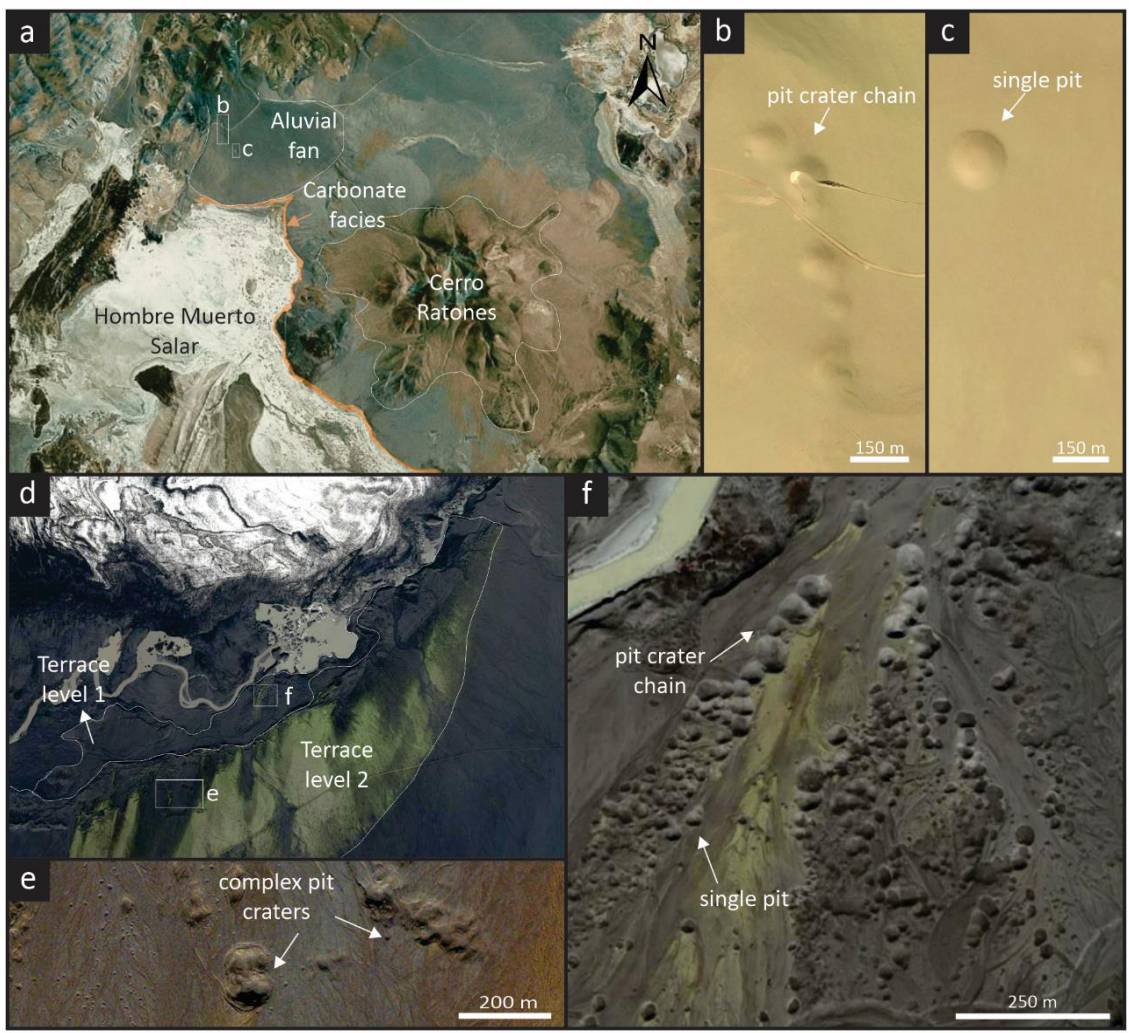

Figure 11: Terrestrial analogues of pit craters. a) Regional context of the alluvial fan with pit craters in Puna, Argentina. The alluvial fan is located in the proximity of Hombre Muerto Salar and near carbonate deposits. b and c) Pit craters chain and single pit identified in the alluvial fan. d) Regional context of the terraces containing pit craters in the Skeiðarárjökull glacier, Iceland. e)

complex pit craters located in the terrace level-2 and f) single pits and pit crater chains located in terrace level-1. Map data 2020-2022 © Maxar Technologies and Landsat/Copernicus.

Despite the Puna being a cold desert with a similar environment to Mars in several ways, pits in Hale are closely related to glacial and periglacial landforms, therefore we also looked for terrestrial glacial environments with pit craters to establish possible analogues. We identified in front of the Skeiðarárjökull glacier (Iceland) fluvial terraces covered by numerous pit crater -like depressions that were previously described as kettle holes, which are typical structures of glacial environments produced by melting of ice blocks buried in the sediment (Fig. 11d-f, Fay, 2002; Magilligan et al., 2002; Russell et al., 2006). The landscape is divided into two levels of terraces with distinct types of pit craters. On the lower level, we observed complex pit craters with circular or subcircular shapes and are 150 and 110 meters long. The area also contains elongated depressions, which resemble a chain of highly deformed pit craters (Fig.11e). In the upper level terrace, we observed single pit craters or chains of pit craters with diameters from 10 to 55 meters, where each individual depression preserves the circular or subcircular geometry (Fig.11f).

The morphological characteristics of the pit craters found in the lower level of terraces could indicate the evolution of these structures. The second level of terraces is in continuous interaction with the water, being an active level that may favor the formation and evolution of the pit craters. In the case of terrace level 1, it is an inactive one and the pit craters represent a less mature stage of their formation. As previously discussed, it is observed that in the Hale Crater area, complex pits are stratigraphically above the first level of aprons, while in the case of single pits are mostly close to gullies channels. Based on this observation, we could suggest that complex and pit crater chains are more mature stages in the formation of these depressions. This observation is also supported by the fact that pit crater chains are formed by the partial coalescence of single pits.

### 5.3. Origin and Evolution of pits

The surface of Mars is covered by different kinds of circular depressions developed by a variety of processes. In the previous sections we described the pit craters located in the northeastern region of Hale Crater. To discuss the origin of the pit craters identified in Hale we must consider the geologic context in which they are found. Typically, pit craters are associated with some type of collapse mainly linked to volcanic, extensional, karstic processes or volatiles (Table 1). However, there is no evidence of volcanic or extensional activity in this case. Although it is possible to establish an analogy with a dissolution process such as that of the Puna, mainly due to the morphological similarity.

Post-impact degradation of impact craters consists of the degradation or complete loss of the raised rim, smoothing on the ejecta blanket and filling in of impact crater bowls (Baker and Carter 2019a, 2019b; Berman et al., 2015; Levy et al., 2009; Mangold 2003; Willmes et al., 2012). Here we identified some depression with a partial elevated rim and no ejecta layer related that could be interpreted as degraded impact craters. This kind of rimless depressions are also characteristic of cratering on icy substrates, such as lobate debris aprons (LDA) or lineated valley fill (LVF), are rapidly degraded due to the sublimation of the ejected material (e.g. Kress and Head, 2008; Levy et al., 2009; Willmes et al., 2012).

Although some single pits might also be interpreted as degraded impact craters, the conical shape of most of them, the close connection with other types of pits lead us to conclude that they are collapse depressions rather than impact craters. In addition, primary impact craters present a random distribution (e. g. Robbins and Hynek, 2014) and the two performed statistical analysis confirmed that the depressions studied here have a non-random distribution. The clustered distribution of the depression studied here rules out a primary impact cratering origin, yet does not rule out the possibility for secondary impact cratering (e. g. Hartmann et al., 2008; Michael et al., 2012; Robbins et al., 2014). In Arcadia Planitia secondary impact craters expanded by sublimation processes are characterized by a central depression with a shallower extension and the loss of the crater rim (Dundas et al., 2015; Viola et al., 2015; Hibbard et al., 2021). At Hale, however, none of the morphologies described for pit craters have a central depression. Certainly, the difference between a degraded impact crater and a single bowl-shaped pit is not easy to discern. Nevertheless, the suite of different morphologies observed, the overlapping of different types of pit craters (Fig. 4f) lead us to interpret these depressions as collapse features, rather than as impact-originated. Furthermore, the depressions with an elevated partial rim do not have a random distribution so we cannot classify them as primary impact craters, although they could be considered as background secondaries (Robbins and Hynek, 2014). A similar approach can be adopted for pit craters; we cannot rule out that some of them may have been originally an impact crater from background secondaries that were later degraded.

We also rule out an aeolian origin for these features, not only because of the assemblage of surrounding volatile-related features, but also because of the lack of any other aeolian features (e.g. TARs, Yardangs) in the area. While in the interior of Hale Crater several dune fields can be identified, no evidence of aeolian erosion can be found in the study area. In addition, if an aeolian origin were to be considered, it would be expected that pits are aligned and elongated in the same direction indicating the wind direction (Ward et al., 1985). The only depressions that show a slightly linear pattern (Fig. 4e) have a partial rim and do not show any elongation, and are concentrated only in the southern area.

The analysis of terrestrial analogues reveals that the pit craters studied in Hale show morphological similarities with pits from arid and periglacial environments on Earth. The Puna pits represent an analog that is only valid if dissolution processes of subsuperficial material are considered. Therefore, we consider that the analysis of new hyperspectral images is required to determine the mineralogy present in the area.

The Icelandic depressions show a relation between the formation of the pit craters, their morphology and the availability of water and the duration of the water-sediment interactions. We consider that Hale pit craters are associated with processes involving the availability of volatiles in the sediment ($CO_2$ and/or $H_2O$ ices). Although recently gullies are associated with $CO_2$ mechanism (e. g. Conway et al., 2019; Diniega et al., 2021; Dundas et al., 2022), several authors suggested in the past that gullies might be formed by mass wasting caused by groundwater sapping (e.g. Heldmann et al., 2007; Heldmann and Mellon, 2004; Malin and Edgett, 2000; Rodriguez et al., 2016). In both cases pit craters could be related with the subsurface removal of material. In one case would be the removal by ice sublimation and in the other the removal by processes similar to sapping.

Based on the geologic context of described pit craters, we propose that they are closely related with volatile processes. Additionally, the clustered distribution of the pit craters on Hale is also consistent with the distribution of periglacial morphologies such as ice mounds (Bruno, 2004; Bruno et al., 2006). There are a variety of landforms related to ice sublimation processes at or near the surface at Martian mid-latitudes, one example is the scalloped depressions of Utopia Planitia and Peneus and Amphitrites Paterae region (e.g. Dundas 2017; Dundas et al., 2015; Lefort et al., 2009; Lefort et al., 2010; Séjourné et al., 2012; Soare et al., 2007; Viola et al., 2018; Ulrich et al., 2012). These are circular to subcircular shallow depressions of hundred meter to kilometer diameters and are associated with polygons. The origin of these depressions is explained by the sublimation of an ice-rich mantling deposits and display a marked north-south asymmetry caused by the changes in insolation in equator/pole facing slopes. Although scalloped depressions are associated with elongated and pit crater chains, these pits are found aligned with polygons and cracks (Lefort et al., 2010; Morgenstern et al., 2007). The comparison of the pit craters studied here with the scalloped depressions results in several important differences. First, the complex pits do not exhibit differences in the equator/pole facing slope and are much smaller than the scalloped depression. Moreover, the profile in Figure 9c shows that the steepest wall is equatorial facing in contrast to what is expected for scalloped depressions. Also, neither the elongated or the pit crater chains present an association with cracks and there is no polygon terrain in the study area. Even though isolated scalloped depressions have been described in the Argyre Basin (Zanetti et al., 2010) we can not relate the depressions described here with those. However, although the pit craters described here are morphologically different to scalloped depressions, they might share a similar origin with collapse features due to volatile/soil interaction. If we considered that some pit craters might be originated by an initial impact craters, post-impact sublimation could have modified the initially crater impact morphology. Moreover, some authors have postulated that scalloped depressions might start as pit craters (Lefort et al., 2010; Morgenstern et al., 2007). Further work might be needed in order to understand why some pits end into scalloped depressions and others just remain as the complex pits described here. Morgenstern et al. (2007) proposed that scalloped depression formation involves the subaerial deposition of volatiles and subsequent degradation of the ice-rich soil. We suggest that the pit craters in this particular location can be explained by this model. In this scenario, the deposition of volatiles must occur along with deposition of sediment that covers the volatile material to prevent a rapid sublimation. In the area there is evidence of multiple depositional episodes: an extensive mantling that covers a wide part of the area and several levels of lobular deposits. However, the pit craters also could be the result of the degradation of the remaining permafrost from the periglacial conditions associated with the moraines-like ridges. In addition, the presence of impact craters with a partial rim similar to the ones recognized over ice-rich substrates (Baker and Carter 2019a, 2019b; Berman et al., 2015; Levy et al., 2009; Mangold 2003; Willmes et al., 2012) is also an evidence of an icy mantling undergoing sublimation processes.

The presence of subsurface ground ice in mid-latitudes of Mars is strongly supported by modeling the stability of subsurface ice (e.g. Mustard et al. 2001; Milliken et al. 2003; Schorghofer and Forget, 2012; Boynton et al. 2002; Feldman et al. 2004). The northern internal wall of Hale Crater exhibits widespread evidence of ice-related landforms, which is consistent with poleward orientation. The presence of pits with a volatile origin described here, at a different slope orientation, shows the complexity of these features. Several variables such as soil thickness, topography, ice distribution or interaction with other processes (gullies) must be taken into account in order to understand these structures.

The analysis of different pit morphologies can provide not only information about the characteristic features of each type of pit craters, it can also give clues about their history. The recognition of simple and some more complex pit craters morphologies, might suggest a progressive modification or evolution of the pits, an evolution pattern from single pits to pit crater chains and complex pits. In this scenario one of the possible origins of the elongated pits might be related to the deformation of the single pits. Although initially this could be associated with particular slope conditions, it is essential to observe a higher number of specimens to reach a conclusion about the slope conditions and the formation of elongated pit craters. Further analysis and modeling are required in order to determine if the different morphologies are in fact different stages in the pit crater formation.

## 6. Conclusion

In our morphological analysis of the pit craters we differentiated 4 types of morphologies (single pits, elongated, chains and complex). Our results showed that there is no specific slope gradient or spectral surface type that would favor the formation of a specific type of pit crater. However, taking into account this new classification, the analysis of more pit craters could provide more robust statistical results that may help to understand these features.

Numerous processes involving the collapse of surface material due to underlying volumetric reduction have been proposed to generate pit craters. In this work we showed that pit craters found in Hale crater are not related to volcanism or tectonic features, but they are in a periglacial environment (section 3). We hypothesized three possible scenarios: a) a dissolution process, b) groundwater sapping, and c) sublimation. Considering those, and the spectral analysis (section 4), we looked for terrestrial analogs for the first two mechanisms (section 5). A better understanding of the selected terrestrial analogues might help to define the origin of martian landforms. Although we could not propose a single clear formation mechanism for the pits in Hale crater, our research points to processes related to volatile availability, most likely sublimation processes, especially considering their association with periglacial features and degraded secondaries impact craters.

Moreover, if we assume that pit craters are related with sublimation processes, the formation and evolution of the pit craters would be conditioned by the distribution and availability of volatiles and climatic conditions. Although insolation is a significant factor, no significant variability in the aspect of pit craters is observed in the study region. Future research will study more closely the relation of insolation parameters and pit locations. In addition, the identification of some depressions that could be possible degraded secondary impact craters is also consistent with this scenario.

We showed that pits are formed before and after gullies and their aprons. Most of the gullies in the area form only within the 'mantling deposits' corresponding to type A2 (Aston et al., 2011). According to the authors, this type of gully is linked with ice-rich deposits. This stratigraphic relation shows a close connection between the two and points to a volatile linked origin for the pits.

For future work we plan to study in depth the thermokarst processes that could be associated with the formation of these depressions, for example by performing experiments of subsurface ice sublimation in a vacuum chamber. In this way, understanding the mechanisms that lead to ground collapse and destabilization could be helpful for the planning of future missions to Mars, for the locomotion of rovers and avoiding rover hazard. Moreover, as mentioned by Viola and McEwen (2018) high-resolution studies are interesting with regards to potentially in situ resource utilization (ISRU) considering that "thermokarst as one indicator of the presence of excess ice".

## Data Availability

The CTX mosaic beta version is described by Dickson et al. (2018) and can be downloaded at http://murray-lab.caltech. The HiRISE images were collected thanks to the open source repository at NASA/JPL/University of Arizona (https://www.uahirise.org/). HiRISE images and CRISM cubes were downloaded from https://ode.rsl.wustl.edu/mars/indexProductSearch.aspx. The analysis of the CRISM hyperspectral images was possible thanks to the open access workshops by Scott Murchie and Frank Seelos and the CRISM Analysis Toolkit (CAT), both available at https://pds-geosciences.wustl.edu/missions/mro/crism.htm. The vector files of the geomorphic map can be accessed through https://doi.org/10.5281/zenodo.7023987.

## Acknowledgments

This is publication number #### from IDEAN. This work was possible thanks to funds of PICT 3697/01796. We also acknowledge the support from the GMAP (Geologic Mapping of Planetary bodies) activity of Europlanet, from the European Union's Horizon 2020 research and innovation programme under grant agreement N° 871149.